# Preference or opportunity? Why do we find more friendship segregation in more heterogeneous schools?[1]

Andreas Flache, Tobias Stark

*Department of Sociology, University of Groningen, a.flache@rug.nl, t.h.stark@rug.nl

**Abstract**

Recently, scholars increasingly use exponential random graph models (*p\**) of friendship network data to disentangle effects of ethnic school heterogeneity on homophily preferences from effects on opportunity for same ethnicity friendships. This research suggests that heterogeneity increases homophily preferences. We argue that this may be a misleading interpretation of *p\** coefficients. If students wish to avoid having no friends at all, then minority students may appear to be willing to integrate more with the majority in a more homogeneous school, even if the preference for having same ethnicity friends has the same strength in all schools. Following Snijder's stochastic actor oriented model, we use a random utility model of network formation that allows direct specification of the preferences that drive the network choices of simulated actors. We conduct computational experiments to generate networks for different school compositions but with the same parameter for a preference for homophily. We estimate with a *p\** model the coefficients for the effect of same ethnicity on (simulated) friendship choice. The tests confirm that *p\** coefficients for the effect of same ethnicity are much larger in populations with a large minority, although actual homophily preferences are the same across all simulated school compositions.

---

[1] This research has been supported by the Netherlands Organisation for Scientific Research, NWO (VIDI Grant 452-04-351). Our work has benefited from stimulating discussions with Christian Steglich, Marijtje van Duin, Michael Mäs, Tom Snijders, Károly Takács, and other members of the discussion group on norms and networks at the Department of Sociology of the University of Groningen.

**1 Introduction**

Racial school segregation remains a persistent topic of both scientific and public debate in the United States (e.g. Mouw & Entwisle, 2006; Quillian & Campbell, 2003; Moody, 2001) and increasingly appears on the agenda in many Western European Countries, where it is driven by a consistent trend towards higher levels of ethnic diversity. For example, in the four biggest cities in the Netherlands about half of all primary schools are considered "black schools", which according to the definition of the Central Statistical Office are schools with more than 50% students with non-western ethnic background, a concentration that far exceeds the national average of 14% non western pupils (CBS, 2007). Researchers and politicians alike are concerned about the negative consequences of school segregation both for inequality between ethnicities in school achievement and for the social integration of society (e.g. Moody, 2001 and Gijsberts & Dagevos, 2005, on this discussion in the U.S. and the Netherlands, respectively). This has lead on both sides of the Atlantic to a range of policies directed at actively fostering mixing of ethnic groups in schools, such as race-based extracurricular mixing or racial busing in the U.S., or selective school admission of minority students in the Netherlands.

Empirical research finds that the mixing of ethnicities in schools may often fail to increase actual interethnic integration, particularly in terms of the formation of friendships across ethnic categories. Based on an analysis of friendship network data from the National Longitudinal Study of Adolescent Health (Add Health) both Moody (2001) and Mouw & Entwisle (2006) conclude that the probability of interracial dyads being friends decreases in more diverse schools and Lubbers (2003) finds at least no positive effect for the Netherlands. This is particularly striking because, as Feld and Carter (1998) have shown, the actual opportunity for forming dyads between majority and minority is larger in more heterogeneous schools. That is, the larger the minority groups are relative to the majority, the larger is the proportion of all theoretically possible dyadic friendship relations across ethnic or racial categories.

Scholars point to preferences for homophily (McPherson, Smith-Lovin & Cook., 2001; Lazarsfeld & Merton, 1954) in friendship selection as the key mechanism that drives higher segregation in more diverse schools. Homophily, or the tendency for "birds of a feather to flock together", may operate both through effects of school heterogeneity on preference and on opportunity. Moody describes the opportunity effect as follows: "in a school where there are many minority students, minority students may be able to find their desired number of friends within the minority friendship pool." (2001, 681). This would result in higher levels of friendship segregation than in a school with a smaller minority, even if students prefer same ethnicity friendships to the same extent. But following previous work on ethnic threat and competition (e.g. Blalock, 1967), Moody also expects that larger minorities may increase the salience of ethnic identity, resulting in stronger preferences for ingroup choice (see also Lubbers, 2003).

Both the scientific explanation of friendship segregation in heterogeneous schools and the design of school integration policies require that we know how much of the effect of school heterogeneity can be attributed to the opportunity mechanism, and how much to direct effects of minority size on homophily preferences. Recently, students of school network data have begun to use exponential random graph, or *p\**, models (cf. Wasserman and Pattison, 1996; for a recent overview see Robins et al, 2007), as a statistical tool to distinguish opportunity effects from preference effects. As Quillian and Campbell (2003, 548) explain: "The first step to estimating the *p\** model is to create dyads that represent all possible friendship pairs within a school. […] This data structure implicitly controls for opportunities for contact among persons of different race in the school, as interracial dyads are represented in proportion to the racial diversity of the student body." Beyond that, the *p\** model is designed to control for the statistical interdependence of various possible characteristics of the network. If, for example, actors prefer same sex friends but are indifferent with regard to the ethnicity of their friends, then we may find segregation as a spurious result when ethnicity and sex happen to be correlated. The *p\** models allows to statistically disentangle such simultaneous effects of different properties of actors, dyads or the network structure as a whole that may affect network choice.

Researchers increasingly use *p\** modeling of friendship network data to disentangle effects of school composition on preferences for same ethnicity friendships from effects driven by opportunity (e.g. Mouw & Entwisle, 2006; Lubbers, 2003; Moody, 2001). For example, Mouw and Entwisle (2006) follow Moody's (2001) preference interpretation of the same-race coefficients obtained from a *p\** analysis. "Moody finds that higher racial diversity decreases the probability of crossrace friendship for any given pair of potential friends of different races. This suggests the existence of 'school climate' effects as a result of heightened racial awareness or tension in racially diverse schools." (2006, p. 398).

We argue in this paper that it may be misleading to interpret effects of school heterogeneity on the same ethnicity coefficients as effects on students' preferences. Exponential random graph models are models of cross sectional network data which – by definition – give no information about the process that brought the structural properties of the network about. But if we find in one network a larger effect of same ethnicity on friendship choice than in another, different causal mechanisms might have brought this difference about. One possibility is that actors have indeed a stronger preference in one network. But another is that they prefer in both networks friends of their own ethnicity to the same extent, but they also have a baseline preference for having any sort of friend rather than no friend at all. This suggests that if minority members can not easily come into contact with other minority members – as in a school with a small minority - they rather would establish a friendship with a member of the other group than have no friendship at all. This may entail biased estimates of the preference for same-race friendships. The same-race coefficient of a $p^*$ model is derived from the odds of a same ethnicity dyad to be a friendship divided by the odds of a ethnically diverse dyad to be a friendship, controlled for other simultaneous effects on the odds ratio of friendship vs. non-friendship. If in a school with a smaller minority, the odds for ethnically diverse dyads to be a friendship are larger than in a school with a larger minority, this can result in a higher coefficient for the same race effect in the school with a larger minority, although students prefer same race friendships to the same extent in both schools.

Furthermore, following theories of bounded rationality (Simon, 1957), we may assume that in their search for friends, students make their decisions myopically. Moreover, people can be assumed to have only a limited capacity for entertaining friendships (cf Zeggelink, 1995). In a school with a small minority, minority students may in the initial phase of friendship formation establish many cross ethnicity ties even if they have a preference for same group friends, due to myopia. But once these ties are established, the students are less likely to continue the search and tend to hold on to these ties, even if they later encounter potential friends from their own group. The result may be a network in which we find not many more same ethnicity ties than we would expect by chance. Hence, if we analyzed this network with a *p\** model we might conclude that students have only a weak homophily preference in a small minority school. But the same students would in a school with a larger majority form a network that would look much more segregated and for which we would obtain much higher coefficients for a homophily "preference" from an analysis with a *p\** model.

To test the logical consistency of our reasoning, we developed a formal theoretical model of friendship choices that explicitly models the preferences of actors and the decision mechanism by which they aim to maximize these preferences within the given opportunity structure. We use computational experiments to derive from this model the resulting dynamics and the emergent structure of the network, given the assumptions about preferences and school composition that we specified. Robins et al (2005) have used a similar approach with a simulation model that is directly based on the *p\** model, but this model can not be seen as an explicit implementation of individual level choice processes. We follow the stochastic actor oriented model proposed by Tom Snijders (1996, 2001, 2005). This model assumes that actors choose in a repeated search process myopically the network ties that maximize their utility given the current state of the network, their (theoretically specified) preferences for certain characteristics of the network and random influences.

We develop a theoretical and computational model of network formation that allows us to directly specify the preferences that drive the network choices of our simulated actors. Then, we use the model to generate networks for different school compositions but with the same parameter for a preference for homophily. We use the simulated networks to estimate with a *p\** model the coefficients for the effect of same ethnicity on (simulated) friendship choice. This allows testing whether p\* coefficients can be affected by school composition even if actual preferences are not.

In section 2, the model is elaborated. Section 3 describes results. Section 4 contains a discussion of our finding and our conclusions.

**2 Model**

The model is closely based on the so-called stochastic actor-oriented models proposed by Snijders (1996, 2001, 2005) and implemented in the SIENA software (Snijders et al, 2007). Stochastic actor-oriented models are primarily developed for statistical modeling of longitudinal empirical data on both networks and actors' attributes. However, an integral component of the statistical method is a theoretical stochastic simulation model. This model derives the expected evolution of both the network relations (e.g. friendships) and changeable actor attributes (e.g. opinions) from theoretically specified assumptions about the underlying preferences of actors and the decision mechanism by which they aim to optimize their choices. The theoretical model underlying SIENA can also be considered as an agent based computational model (cf. Macy and Willer 2002). Accordingly, we use in the following the terms 'actor' and 'agent' interchangeably and refer sometimes also to 'students' and 'friendships' when talking about our artificial school population.

We re-implemented in a general purpose programming language (DELPHI) a somewhat simplified version of the simulation algorithm of SIENA in which we could include all of the assumptions that we want to specify. For our purposes, it suffices in particular to only model choices with regard to network relations, because we are interested in only one actor attribute that we assume to not be changeable, actors' ethnic category. For simplicity, we distinguish only two categories 'minority member' and majority member, expressed by the value of the attribute $a_i$ for actor $i$ (1=minority, 0=majority). Moreover, we included for the time being only a limited set of preferences for network structural characteristics, notably preferences with regard to the desired outdegree, the reciprocation of ties and same ethnicity of friends.

Following SIENA, we model actors' network choices as the result of myopic maximization of the "random utility" of making or breaking a particular network tie. That is: given the assumptions that we make about actors' preferences, adding or deleting a certain tie in the personal network yields for the focal actor a certain utility that he derives from the resulting new network. For example, if we assume that actors only have preferences for reciprocated ties above unreciprocated ties and ties to similar friends above those with dissimilar friends, then the highest utility and actor $i$ can obtain with one single change in her network would be derived from establishing a tie to a similar other *j* who already has chosen *i* as a friend. However, following the approach of random utility models, SIENA models also assume that there is a random component of the utility of an action. It is assumed that actors maximize the actual utility of an action, which is a sum of the systematic component and the random component. This implies that the principle of utility maximization yields a multinomial probabilistic choice: the higher the utility of a certain network choice according to our model of actors' preferences, the higher the probability that this option will be selected, but for all choices there is always a positive probability (see Snijders, 2001).

More precisely, the model assumes that the network changes in small steps: at each given moment, only one tie variable $Y_{ij}$ can change, where $Y_{ij}$ is a binary variable indicating by the values 1 or 0 whether or not there is a tie from *i* to *j*. In every iteration of the simulation, one actor is selected randomly and gets the opportunity to change one of his tie variables. When *i* gets this opportunity he or she selects the best possible change according to a myopic stochastic optimization procedure. The utility an actor *i* derives from a network state *y* is expressed by the so-called objective function defined as $\Sigma_k \beta_k s_{ki}(y)$ where the $s_{ki}(y)$ are network statistics and the $\beta_k$ are parameters expressing the strength of each "effect" or utility argument $s_{ki}(y)$. An example of a network statistic $s_{ki}(y)$ is the out-degree of *i*, another one is the number of *i*'s reciprocated ties, or the number of actors to whom *i* is tied and who have the same ethnic category than *i*.

When *i* has been selected for an option to change one of her ties, the selection decision is conducted as follows. In a first step, for every possible change of *i*'s current network the objective function is calculated for the network that would result from the change. We also include the possibility of a non-change. This results in a vector **o** of *N* values of the objective function for *N* different possible networks (with *N* being the number of actors in the network). The elements 1..*N* express the values of the objective function that would result if *i* would change the value of the current tie $y_{ij}$ into it's opposite 1-$y_{ij}$, with one exception. The model excludes the possibility of a tie to self, thus we use element *i* of the vector to hold the value of the objective function if the current network is not changed.

We use the following utility arguments that together with the corresponding parameters $\beta_k$ constitute the objective function:

Outdegree:
$$s_{0i}(y) = \sum_{j \neq i} y_{ij} \quad (1)$$

Number of reciprocated ties:
$$s_{1i}(y) = \sum_{j \neq i} y_{ij} y_{ji} \quad (2)$$

Number same ethnicity friends:
$$s_{2i}(y) = \sum_{j \neq i} y_{ij}(a_i a_j + (1-a_i)(1-a_j)) \quad (3)$$

For the number of reciprocated ties and the number of same ethnicity friends, we follow the standard approach of SIENA and assume that they have a linear effect on the objective function. That is, any additional reciprocated tie or any additional similar friends increases (or decreases) the objective function by the values of the corresponding parameters $\beta_1$ and $\beta_2$. For the effect of outdegree we choose a different assumption that to our knowledge can not be modeled with the currently available network effects in SIENA. Following previous models of network choices, we assume that people have a basic desire for having friends, but also a limited capacity for relationships (c.f. Zeggelink 1995). More precisely, we assume that for agents who have a number of friends that falls below some threshold level, it is desirable to increase their personal network, but if the number of friends has reached some limit, additional friends are no longer desirable. More technically, for agents with a low outdegree, the marginal utility of an additional tie exceeds the marginal costs, regardless of the characteristics of the agent to whom the additional tie is established. But as outdegree increases, marginal costs increase likewise such that at some point additional friends reduce the value of the objective function even for the most attractive type of friendship added. Technically, we model this assumption with a parameter for the outdegree utility argument that declines linearly in outdegree, as expressed by equation (4).

$$\beta_0(s_{0i}(y)) = \beta_{0,0} + \beta_{0,1} s_{0i}(y) \quad (4)$$

In equation (4), the intercept term $\beta_{0,0}$ models the effect of an additional tie on utility for an empty ego-network. We assume that this parameter is positive. The slope term $\beta_{0,1}$ indicates how much the utility of an additional tie changes when the outdegree increases. To model declining marginal utility, we assume this parameter to be negative.[2]

The first step in modeling network decisions is to compute for every possible change of a single tie (including non-change) the corresponding values of the objective function after the change. The second step is to derive from these values the choice probabilities. Following the conventional form of random utility models, the probability that choice option *k* is selected is obtained from the exponentiation of the corresponding value of the objective function divided by a normalization factor that assures that the sum of the probabilities across all possible choices equals one. Technically, the probability for choice option *m* is a function $p_{im}(o_m)$ of the value of the objective function after the corresponding choice, as expressed in $o_m$ computed in the first step, formalized in equation (5).

$$p_{ik}(o_m) = \frac{\exp(o_m)}{\sum_l \exp(o_l)} \qquad (5)$$

The third step is to determine the actual choice. For this, a pseudo random experiment is conducted that selects exactly one of the possible choice, with success probabilities for the outcome $m \in 1..N$ determined by the probabilities $p_{im}(o_m)$. Once the decision *m* has been selected, the network is updated accordingly. The dynamics of the model result from repeated application of this decision mechanism in a sequence of iterations.

*Outcome measures*

To show how model assumptions affect the dynamics and structure of the simulated networks, we calculate and display four outcome measures. The program displays the density of the simulated network (proportion of dyads for which $y_{ijt}=1$). Our first measure of ethnic segregation is the proportion of all friendship ties that are between friends who differ in ethnicity (*pDiss*). If this measure adopts its theoretical minimum of zero, then none of the ties in the network is a cross ethnicity tie. The theoretical maximum of one indicates that all friendship ties are cross ethnicity. This measure is useful to assess how much cross ethnicity relations there are overall relative to the density of the network. However, it does not allow us to compare segregation levels between populations with different composition, because the maximum of the proportion of cross ethnicity ties depends on the relative size of the minority. We adopt from Moody (2001) two different measures that do not have this undesirable property. Moody distinguishes between indicators of "gross segregation" and "net segregation". Gross segregation describes the extent to which network ties are more likely between actors with the same ethnicity than between ethnically dissimilar actors. Moody measures this with the odds ratio $\alpha$ of a same-ethnicity by friendship cross-tabulation. Let A denote the number of friendship ties between members of the same ethnicity, and C the number of (ordered) same ethnicity pairs without a friendship tie. Then, the odds that there is a friendship $i \rightarrow j$ if $i$ and $j$ have the same ethnicity are A/C. Correspondingly, we obtain the odds that there is a friendship between actors with different ethnicity as B/D where B is the number of friendships $i \rightarrow j$ between ethnically dissimilar i's and j's, and D is the corresponding number of non-friendships. The odds ratio $\alpha$ is then obtained as AD/BC.

---

[2] The assumption that agents with an empty ego network do not resist friendships with dissimilar others may not be plausible for all contexts, but it seems very adequate for our purposes. We focus on only one attribute in which agents can be dissimilar, ethnic category. And we know from school research on interethnic friendship choices that substantial levels of interethnic friendships occur also in schools where members of the minority have little chance to find same-ethnicity friends (e.g. Quillian & Campbell, 2003). Obviously, the lack of potential same ethnicity friends does not prevent them from forming relationships.

Net segregation is the level of segregation in friendship choices that we observe net of other characteristics of potential friends, the dyad, or the network that may affect the choice. We obtain for all conditions that we simulate the net segregation level by estimating a *p\** model of the network in the final round of the simulation. To assure that the stochastic process that we simulate has approached a stationary distribution, we run the simulations for an average of 100 iterations per agent. Inspection of model dynamics across multiple independent replications showed that these outcome measures had stabilized in the final round of the simulation where we estimated net segregation.

We use for estimating the *p\** model the implementation of the estimation algorithm for exponential random graph models (ERGM) in the SIENA software (cf. Snijders et al 2007). The *p\** model that we estimated follows the conventional form used in studies such as Moody (2001), Lubbers (2003) or Mouw and Entwisle (2006), but is much simpler than the models used by these authors, because it contains only the effects of outdegree (density), reciprocity and same ethnicity. We know that only the corresponding preferences can affect the network choices of our simulated agents, because we have programmed our agents accordingly. Following Moody (2001), we represent the *p\** model in the form a of a logistic regression model given by (6).

$$Log\left(\frac{p(Y_{ij}=1)}{p(Y_{ij}=0)}\right) = b_0 + b_1 \text{reciprocal} + b_2 \text{same ethnicity} + e_{ij} \quad (6)$$

The parameter $b_0$ in this model corresponds to the effect of network density on the likelihood of tie $y_{ij}$ to be present. A negative density effect shows that higher density of the rest of the network makes a tie $y_{ij}$ less likely to exist (ceteris paribus). We are aware that the constant density effect in the *p\** model does not match the theoretical specification of a non-linear outdegree effect in our simulation model. However, we want to know which conclusions researchers would draw from an analysis of network data generated with our theoretically specified model of preferences, if they apply a *p\** model with the conventional form. Parameter $b_1$ of the *p\** model expresses the estimated effect of reciprocation on the likelihood of the tie $y_{ij}$ to exist, and $b_2$ captures the corresponding effect of same ethnicity of *i* and *j*. The term $e_{ij}$ is the error at the dyadic level. We take the parameter $b_2$ as measure of net segregation. To be precise, we report for each of the conditions that we simulated average coefficients of the *p\** model that we obtained from estimations of the coefficients for each of a number of different networks generated by independent replications of the simulation. Then, we employed the multilevel extension of the SIENA software described in Snijders et al (2007, chapter 14), which implements a meta-analysis of the coefficient estimates of multiple applications of the same *p\** model. This meta-analysis models the distribution of the coefficient estimates across replications, controlling for differences in the standard errors of the estimates between replications (cf. Lubbers, 2003).

## 3. Results

We conducted a range of computational experiments with different assumptions about actors' preferences with regard to homophily. In each of the experiments, we varied the proportion of minority members from 0.1 to 0.5 in steps of 0.1 and obtained the outcome measures for each condition. We report here results for a population size of *N*=50. Experiments with population sizes of 20, 30 and 40 yielded the same qualitative results. We assumed throughout the experiments reported here that the process starts from an empty network ($\forall ij : y_{ij,t=0} = 0$) and continues for 5000 iterations per replication (100 per agent). For control, we also tested effects of using an initial random network with a density of 30% rather than an empty network. We found no qualitative differences in the results. We did not vary the assumptions about preferences for outdegree or reciprocity, because the effects of these parameters are not our theoretical focus. We assumed for all experiments that the preference parameter for outdegree is a function with a negative slope of $\beta_{0,1} = -0.5$ but a positive intercept of $\beta_{0,0} = 10$. With this, the contribution of an additional tie to the objective function turns negative when the outdegree exceeds 20 (neglecting reciprocity and homophily effects). We fixed the reciprocity preference parameter of the simulation model to $\beta_1 = 1$.

To establish a baseline, we simulated a model that assumes no homophily. With $\beta_2 = 0$, this model implies that agents are indifferent between forming a tie to a member of the same or another ethnicity, all other characteristics of the tie being equal. Figure 1 charts the average dynamics of density, proportion of dissimilar ties (*pDiss*) and gross segregation ($\alpha$) that this model generates for a minority size of 30% of the population. Notice that density and *pDiss* are naturally constrained to the interval [0..1], while $\alpha$ can take any positive value, but remains in the interval [0..2] for the condition simulated here.

[Figure 1 about here]

Figure 1 shows that consistently with the specification of the preference parameters, the model generates no tendency towards segregation in the baseline condition. The average density of networks increases in approximately the first 500 iterations and stabilizes thereafter to a level of about 25%, corresponding to an average outdegree of about 12.5. The gross segregation level stabilizes quickly close to 1.0, indicating that there is no discernable difference between the odds of a same ethnicity dyad to be a friendship and the odds of a different ethnicity dyad. The proportion of dissimilar ties is almost from the first iteration on stable at about 0.425, close to the likelihood that with a minority of 30% a randomly chosen dyad is a cross ethnicity dyad.

Next, we wanted to test how the size of the minority affects segregation levels according to the baseline model. Feld and Carter (1998) showed that the number of cross ethnicity ties should be largest in schools with the largest possible minority, if students are indifferent to ethnicity in their friendship choices. At the same time, the model should generate no effects on gross segregation. Figure 2 reports the results based on outcome measures obtained after 5000 iterations, averaged across 100 independent replications per condition.

[Figure 2 about here]

Figure 2 confirms that there is no segregation in terms of gross segregation across all minority sizes, if we assume indifference to ethnicity in friendship choices. The figure also shows that the average density of the simulated networks is stable across different minority sizes. The only effect of minority size is that the proportion of ties that are formed between agents with different ethnicity increases in the size of the minority. This is consistent with Feld's and Carters' (1998) argument.

To test effects of a preference for homophily, we conducted a ceteris paribus replication of the experiment for a model in which we assumed a preference for same ethnicity friendships ($\beta_2 = 1$). Figure 3 reports the results for the effect of minority size on segregation. We do not report density in figure 3, because for all minority sizes it differed only marginally from the density that we obtained in the experiment reported in figure 2. Values for $\alpha$ are now substantially larger than in the previous experiment such that we have extended the range of the vertical scale to [0..15]. For better visualization, we report *pDiss* now in %, where values never exceed the range between 0 and 15%.

[Figure 3 about here]

Figure 3 supports the expectation of higher levels of segregation in populations with larger minorities, despite constant preferences across all minority sizes. We find that gross segregation increases from about 2.9 for a minority of size 0.1 to approximately 12.84 when the "minority" is 50% of the population. The observed level of gross segregation is about 4 times higher in a large minority population than in a small minority population, but the actual underlying preferences for forming ingroup friendships are the same in both conditions. Further tests showed that this pattern generalized to other positive values of the homophily preference $\beta_2$, with gross segregation levels increasing stronger in minority size for higher values of $\beta_2$.

Inspection of the proportion of dissimilar ties shows that the effect highlighted by Feld and Carter (1998) now only obtains for very small minorities. For minorities of 20% or larger, a larger size of the minority does no longer increase the number of dissimilar ties. About 12% of all ties are cross ethnicity ties, across all sizes of the minority in this range. Only for very small minorities (10%) we observe less cross ethnicity ties in absolute terms than for larger minorities. When we increased $\beta_2$ further to $\beta_2 = 2.0$, we even found that the number of cross ethnicity ties was highest for the smallest minority of 10%.

Our results so far support the possibility that we may in reality observe substantially more segregation in schools with larger minorities, even if the underlying homophily preference of students were not affected by school composition. But we have not yet analyzed how according to our model ethnic composition of the population affects the measure of net segregation obtained from the *p\** model. If the *p\** model does indeed allow to disentangle composition effects from preference effects in friendship choices, we should see that the estimated coefficients for the effect of same ethnicity on the likelihood of a friendship choice does not change with the composition of the population.

To begin with, we estimated the *p\** model of (6) for the networks generated by the simulation model for the baseline condition ($N=50$, $\beta_{0,0} = 10$, $\beta_{0,1} = -0.5$, $\beta_1 = 1, \beta_2 = 0$). We compared the parameter estimates for the smallest and the largest minority size that we simulated, 10% and 50% respectively. For both minority sizes, parameter estimates and standard errors are obtained from a meta analysis of the *p\** coefficients that we obtained for each of the 100 networks generated per condition. Results are displayed in table 1.

[Table 1 about here]

Table 1 shows that the *p\** model captures adequately the theoretical assumptions about agents' preferences that we specified in the simulation model. The negative and significant density effect (reported in column $\hat{\mu}_\theta^{wls}$) reflects our assumption that for a given outdegree actors tend to refrain from forming new ties. The *p\** analysis identified correctly a significant and positive effect of reciprocity for both minority sizes, and the insignificant coefficients for same ethnicity reflect likewise the assumption of indifference to ethnicity that we made for the baseline condition. Notice that the sizes of the coefficients can not be directly compared to the parameters that we assumed for the simulation model, simply because the two models (*p\** and our SIENA based simulation model) have a different structure.[3]

Next, we replicated the analysis for the networks generated in the conditions assuming homophily for the 100 networks that were generated for each of the two minority sizes of 10% and 50%, respectively. Table 2 reports the results.

[Table 2 about here]

Table 2 confirms our expectation that an analysis of "net segregation" based on a *p\** model can not entirely disentangle effects of the preference for choosing same ethnicity friends from effects of differences in opportunity between schools with small and large minorities. There is no difference between small and large minority populations in the preference parameters of the process model that generated the networks. However, the same ethnicity coefficient obtained from the *p\** models is more than twice as large for minority size 50% than for minority size 10%.

**Discussion and Conclusion**

We conclude that evidence for effects of school composition on homophily coefficients of ERG models can not readily be interpreted as indication of an effect on the underlying homophily preference of students. We have explicitly modeled the preferences that guide students' network choices and the decision mechanism by which choices are made. We compared artificial populations with small vs. large minorities, assuming that in all schools students prefer friends from their own group to the same extent. Consistently with the empirical pattern, our model generates dramatically higher levels of segregation in more heterogeneous schools. This difference remained when we used a *p\** model to obtain "net segregation" levels which - according to recent studies – can be seen as a measure of the homophily preference controlled for composition effects.

---
[3] The very small standard deviation parameter and its never significant Q-statistic indicate that there is no variation in the network statistics between our simulated school samples.

There are some limitations of our work that point to directions for future research. First, neither in our simulation model nor in the $p^*$ analysis did we take into account transitivity effects. But network researchers have recognized for long that transitivity is an important element of social networks (Holland & Leinhardt 1971). Second, we have not addressed the problem of how to obtain more reliable estimates of homophily preferences from school network data. We expect that longitudinal network data in combination with actor oriented statistical modeling (Snijders, 1996, 2001, 2005) may allow to better disentangle preferences and composition effects. We will test this expectation both in future theoretical studies and with analyses of longitudinal school network data that we are currently collecting.

**Figures**

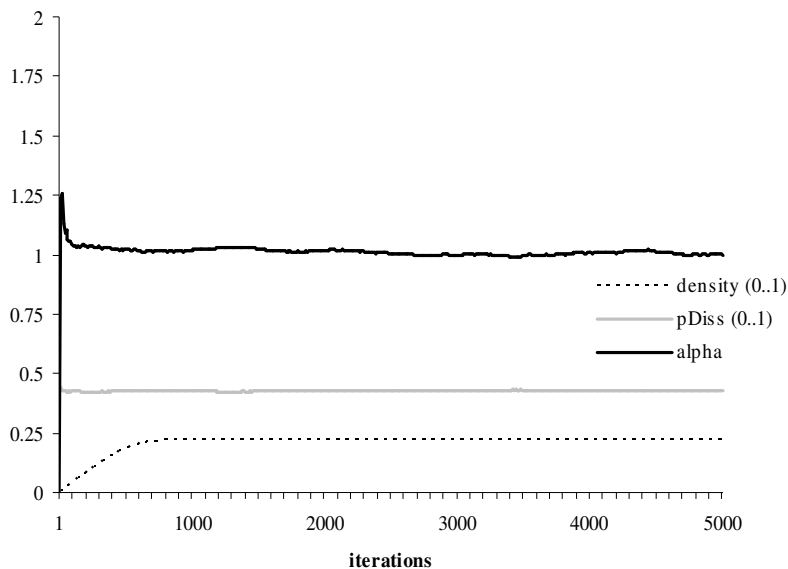

Figure 1. Change of density, proportion of dissimilar ties (*pDiss*) and gross segregation ($\alpha$) in baseline scenario. Averages of 100 independent replications. 30% minority. $N=50, \beta_{0,0} = 10, \beta_{0,1} = -0.5, \beta_1 = 1, \beta_2 = 0$.

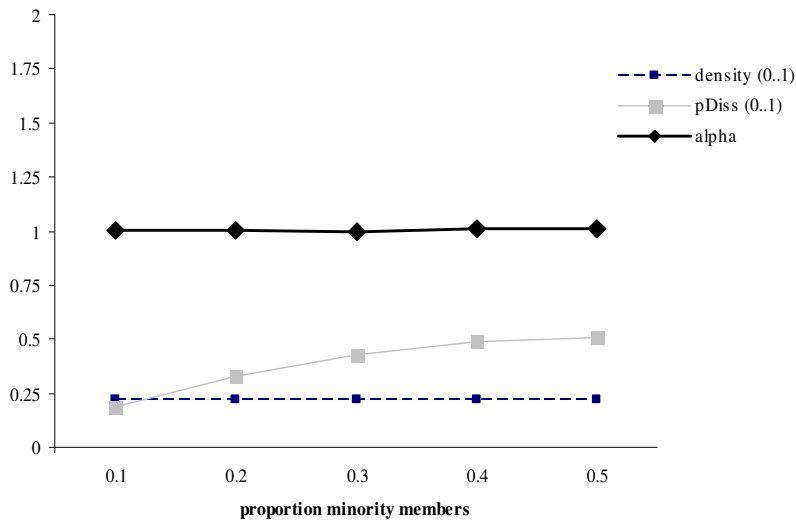

Figure 2. Effect of proportion minority members on network density, proportion dissimilar ties (*pDiss*) and gross segregation ($\alpha$) after 5000 iterations in baseline scenario. Averages of 100 independent replications per condition. $N=50$, $\beta_{0,0} = 10$, $\beta_{0,1} = -0.5$, $\beta_1 = 1, \beta_2 = 0$.

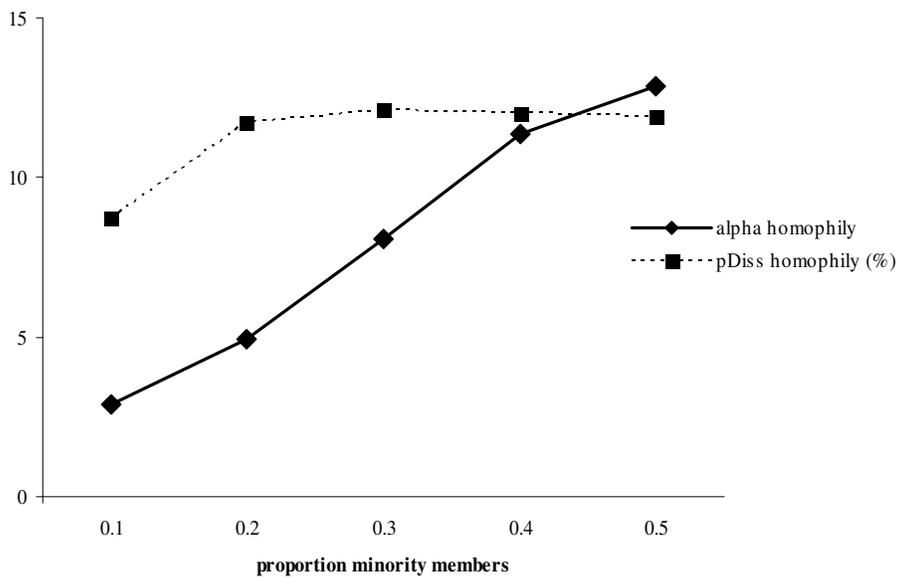

Figure 3. Effect of proportion minority members on gross segregation ($\alpha$) and proportion dissimilar ties after 5000 iterations in scenario with homophily ($\beta_1 = 1, \beta_2 = 1$). Averages of 100 independent replications per condition. $N=50$, $\beta_{0,0} = 10$, $\beta_{0,1} = -0.5$.

## Tables

*Table 1*: Results of $p^*$ meta analysis for baseline condition ($\beta_1 = 1, \beta_2 = 0$).

|  | $T^2$ | $\hat{\mu}_\theta^{wls}$ | (s.e.) | $\sigma_\theta$ | Q |
|---|---|---|---|---|---|
| *Small minority = 0.1* | | | | | |
| - Density ($b_0$) | 59254** | -1.846 | (.008)** | .000 | 61 |
| - Reciprocity ($b_2$) | 14313** | 1.945 | (.017)** | .040 | 94 |
| - Same ethnicity ($b_1$) | 49 | .001 | (.011) | .000 | 48 |
| *Big minority = 0.5* | | | | | |
| - Density ($b_0$) | 48613** | -1.829 | (.008)** | .000 | 89 |
| - Reciprocity ($b_2$) | 15029** | 1.909 | (.017)** | .061 | 112 |
| - Same ethnicity ($b_1$) | 100 | .007 | (.009) | .015 | 99 |

$T^2$: statistic for testing that total effect is zero
$\hat{\mu}_\theta^{wls}$: estimated average effect size and its standard error (s.e.)
$\sigma_\theta$: estimated true between-schools standard deviation of the effect size
Q: statistic for testing that true effect variance is zero
\* p < .01; ** p < .001

*Table 2*: Results of $p^*$ meta analysis for "homophily included" ($\beta_1 = 1, \beta_2 = 1$).

|  | $T^2$ | $\hat{\mu}_\theta^{wls}$ | (s.e.) | $\sigma_\theta$ | Q |
|---|---|---|---|---|---|
| *Small minority = 0.1* | | | | | |
| - Density ($b_0$) | 30339** | -2.4274 | (.014)** | .000 | 46 |
| - Reciprocity ($b_2$) | 15320** | 1.8838 | (.017)** | .079 | 123 |
| - Same ethnicity ($b_1$) | 3117** | .7768 | (.014)** | .000 | 39 |
| *Big minority = 0.5* | | | | | |
| - Density ($b_0$) | 65100** | -3.0633 | (.012)** | .016 | 97 |
| - Reciprocity ($b_2$) | 12355** | 1.9109 | (.017)** | .000 | 95 |
| - Same ethnicity ($b_1$) | 20317** | 1.8954 | (.014)** | .041 | 106 |

Symbols: see footnote table 1; * p < .01; ** p < .001